\documentclass[amsmath,twoside,superscriptaddress,showpacs]{iopart}

\usepackage{iopams}
\usepackage{graphicx}

\def\hmpc{\ {\rm {\it h}^{-1}Mpc}}

\def\hmmpc{\ {\rm {\it h}Mpc^{-1}}}

\def\hmsun{\ {\rm M_\odot/{\it h}}}

\def\la{\langle}
\def\ra{\rangle}

\def\dc{\delta_{\rm c}}
\def\dec{\delta_{\rm ec}}

\def\pmm{P_\delta}
\def\pmh{P_{{\rm h}\delta}}
\def\phh{P_{\rm h}}

\def\fnl{f_{\rm NL}^{\rm X}}
\def\floc{f_{\rm NL}^{\rm loc}}
\def\feq{f_{\rm NL}^{\rm eq}}
\def\ffol{f_{\rm NL}^{\rm fol}}
\def\gloc{g_{\rm NL}^{\rm loc}}

\def\vk{{\bf k}}
\def\vq{{\bf q}}
\def\vx{{\bf x}}
\def\vr{{\bf r}}

\def\TM{{\cal M}_R}

\begin{document}

\submitto{\CQG}

\topmargin-1cm

\title{Primordial non-Gaussianity from the large scale structure}

\author{V Desjacques$^1$, U Seljak$^{1,2}$}
\address{$^1$ Institute for Theoretical Physics, University of Z\"urich, 
Winterthurerstrasse 190, CH-8057 Z\"urich, Switzerland}
\address{$^2$ Physics and Astronomy Department, University of California, 
and Lawrence Berkeley National Laboratory, Berkeley, California 94720, USA}
\eads{\mailto{dvince@physik.uzh.ch}, \mailto{seljak@physik.uzh.ch}}

\date{\today}

\begin{abstract}

Primordial non-Gaussianity is a potentially powerful discriminant of
the physical mechanisms that generated the cosmological fluctuations
observed today. Any detection of non-Gaussianity  would have profound
implications for our understanding of cosmic structure formation.  In
this paper, we review past and current efforts in the search for
primordial non-Gaussianity in the large scale structure of the
Universe.
 
\end{abstract}

\maketitle

\setcounter{footnote}{0}

\section{Introduction}
\label{sec:intro}

In generic inflationary models based on the slow roll of a scalar
field, primordial curvature perturbations are produced by the inflaton
field as it slowly rolls down its potential
\cite{1981JETPL..33..532M,1982PhLB..117..175S,
1982PhLB..115..295H,1982PhRvL..49.1110G}. Most of these scenarios
predict a nearly scale-invariant spectrum of adiabatic curvature
fluctuations, a relatively small amount of gravity waves and tiny
deviations from Gaussianity in the primeval distribution  of curvature
perturbations
\cite{1987PhLB..197...66A,1992PhRvD..46.4232F,1994ApJ...430..447G}.
While the latest measurements of the cosmic microwave background (CMB)
anisotropies favor a slightly red power spectrum
\cite{2009ApJS..180..330K}, no significant detection of a $B$-mode or
of primordial non-Gaussianity (NG) has thus far been reported from CMB
observations.

While the presence of a $B$-mode can only be tested with CMB
measurements \cite{1997PhRvL..78.2054S,1997PhRvL..78.2058K},
primordial deviations from Gaussianity can leave a detectable
signature in the distribution of CMB anisotropies {\it and} in the
large scale structure (LSS) of the Universe. Until recently, it was
widely accepted that measurement of the CMB furnish the  best probe of
primordial non-Gaussianity \cite{2000MNRAS.313..141V}.  However, these
conclusions  did not take into account the scale-dependence of the
galaxy power spectrum and bispectrum arising for primordial NG of the
local $\floc$ type
\cite{2004PhRvD..69j3513S,2008PhRvD..77l3514D}. These theoretical
results, together with rapid developments in observational techniques
that will provide large amount of LSS data, will enable us to
critically confront predictions of non-gaussian models. In particular,
galaxy clustering should provide independent constraints on the
magnitude of primordial non-Gaussianity as competitive as those from
the CMB and in the long run may even give the best constraints.

The purpose of this work is to provide an  overview of the search for
a primordial non-Gaussian signal in the large scale structure. We will
begin by briefly summarizing how non-Gaussianity arises in
inflationary models (\S\ref{sec:models}). Next, we will discuss the
impact of primordial non-Gaussianity on the mass distribution in the
low redshift Universe (\S\ref{sec:matterng}). The main body of this
review is \S\ref{sec:lssprobes}, where we describe in detail an number
of methods exploiting the abundance and clustering properties of
observed tracers of the LSS to constrain the amount of initial
non-Gaussianity.  We conclude with a discussion of present and
forecasted constraints achievable with LSS surveys
(\S\ref{sec:limits}).

\section{Models and observables}
\label{sec:models}

Single-field slow-roll models lead to a very small level of primordial
non-Gaussianity
\cite{1990PhRvD..42.3936S,1992PhRvD..46.4232F,1994ApJ...430..447G}.
This is because they assume i) a single dynamical field (the inflaton)
ii) canonical kinetic energy terms (i.e.  perturbations propagate at
the speed of light) iii) slow roll (i.e. the timescale over which the
inflaton field changes is much larger than the Hubble rate) iv) an
initial adiabatic Bunch-Davis vacuum.  The lowest order statistics
sensitive to non-Gaussian features in  the initial distributions of
scalar perturbations $\Phi(\vx)$  (We  shall adopt the standard CMB
convention in which $\Phi(\vx)$  is the Bardeen's curvature
perturbation in the matter era) is the 3-point function or  bispectrum
$B_\Phi(\vk_1,\vk_2,\vk_3)$, which is a function of any triangle
$\vk_1+\vk_2+\vk_3=0$ (as follows from statistical homogeneity which
we assume throughout this paper). It has been shown that, in the
squeezed limit  $k_3\ll k_1\approx k_2$, the bispectrum of {\it any}
single-field  slow-roll inflationary model asymptotes to the local
shape
\cite{2003JHEP...05..013M,2003NuPhB.667..119A,2004JCAP...10..006C}.
The corresponding nonlinear parameter predicted by these models is
$\floc=\frac{5}{12}\left(1-n_s\right)\approx 0.017$ where $n_s$ is the
tilt or spectral index of the power spectrum $P_\Phi(k)$, which is
accurately measured to be   $n_s\approx 0.960\pm 0.013$
\cite{2009ApJS..180..330K}. Therefore, any robust measurement of
$\floc$ well above this level would thus rule out single-field
slow-roll inflation as defined above.

\subsection{The shape of the primordial bispectrum}

Large, potentially detectable amount of Gaussianity can be produced
when at least one of the assumptions listed above is violated, i.e. by
multiple scalar fields \cite{1997PhRvD..56..535L,2003PhRvD..67b3503L},
nonlinearities in the relation between the primordial scalar
perturbations and  the inflaton field
\cite{1990PhRvD..42.3936S,1994ApJ...430..447G}, interactions of scalar
fields \cite{1993ApJ...403L...1F}, a modified dispersion relation or a
departure from the adiabatic Bunch-Davies ground state
\cite{1997NuPhB.497..479L}. Generation of a large non-Gaussian signal
is also expected at reheating \cite{2004PhRvD..69h3505D} and in the
ekpyrotic scenario  \cite{2008PhRvD..77f3533L}. Each of these physical
mechanisms leaves a distinct signature in the primordial  3-point
function $B_\Phi(\vk_1,\vk_2,\vk_3)$, a measurement of which would
thus provide a wealth of information about the physics generating the
primordial fluctuations. Although the configuration shape of the
primordial bispectrum can be extremely complex in some models, there
are broadly three classes of shape characterizing the local,
equilateral and folded type of primordial non-Gaussianity
\cite{2004JCAP...08..009B,2009PhRvD..80d3510F}. The magnitude of each
template ``X''  is controlled by a dimensionless nonlinear parameter
$\fnl$ which we seek to constrain using CMB or LSS observations.

Any non-Gaussianity generated outside the horizon induces a
three-point function that is peaked on squeezed or collapsed triangles
for realistic values of the scalar spectral index. The resulting
non-Gaussianity depends only on the local value of the Bardeen's
curvature potential and can thus be conveniently parameterized up to
third order by
\cite{1990PhRvD..42.3936S,1994ApJ...430..447G,2000MNRAS.313..141V,
2001PhRvD..63f3002K}
\begin{equation}
\Phi(\vx)=\phi(\vx)+\floc\phi^2(\vx)+\gloc\phi^3(\vx)\;,
\label{eq:philoc}
\end{equation} 
where $\phi(\vx)$ is an isotropic Gaussian random field and $\floc$,
$\gloc$ are dimensionless, phenomenological parameters. Since
curvature perturbations are of magnitude ${\cal O}(10^{-5})$, the
cubic order correction should always be negligibly small compared to
the quadratic one when ${\cal O}(\floc)\sim{\cal O}(\gloc)$. However,
this condition is not satisfied by some multifield inflationary models
such as the curvaton scenario, in which a large $\gloc$ and a small
$\floc$ can be simultaneously produced \cite{2003PhRvD..67b3503L}.
The quadratic term generates the  3-point function at leading order,
\begin{equation}
\label{eq:bphiloc}
B_\Phi^{\rm loc}(\vk_1,\vk_2,\vk_3)=2\floc\left[P_\phi(k_1)P_\phi(k_2)
+\mbox{(cyc.)}\right]\;,
\end{equation}
where (cyc.) denotes all cyclic permutations of the indices and
$P_\phi(k)$ is the power spectrum of the Gaussian part $\phi(\vx)$ of
the Bardeen  potential. The cubic-order terms generates a trispectrum
$T_\Phi(\vk_1,\vk_2,\vk_3,\vk_4)$ at leading order.

Equilateral type of non-Gaussianity, which arises in inflationary
models with higher-derivative operators such as the DBI model, is well
describe by the factorizable form \cite{2006JCAP...05..004C}
\begin{eqnarray}
\label{eq:bphieq}
\fl
B_\Phi^{\rm eq}(\vk_1,\vk_2,\vk_3) = && 6\feq
\Bigl[-\bigl(P_\phi(k_1)P_\phi(k_2)+\mbox{(cyc.)}\bigr)
-2\bigl(P_\phi(k_1)P_\phi(k_2)P_\phi(k_3)\bigr)^{2/3}\Bigr.  \\
&& \quad +  \Bigl. \bigl(P_\phi^{1/3}(k_1)P_\phi^{2/3}(k_2) P_\phi(k_3)
+\mbox{(perm.)}\bigr)\Bigr]\nonumber \;.
\end{eqnarray}
It can be easily checked that the signal is largest in the equilateral
configurations $k_1\approx k_2\approx k_3$, and suppressed in the
squeezed limit $k_3\ll k_1\approx k_2$. Note that, in single-field
slow-roll inflation, the 3-point function is a linear combination of
the local and equilateral shape \cite{2003JHEP...05..013M}.

As a third template, we consider the folded or flattened shape  which
is maximized for $k_2\approx k_3\approx k_1/2$
\cite{2009JCAP...05..018M}
\begin{eqnarray}
\label{eq:bphifol}
\fl
B_\Phi^{\rm fol}(\vk_1,\vk_2,\vk_3) = && 6\ffol
\Bigl[\bigl(P_\phi(k_1)P_\phi(k_2)+\mbox{(cyc.)}\bigr)
+3\bigl(P_\phi(k_1)P_\phi(k_2)P_\phi(k_3)\bigr)^{2/3}\Bigr.  \\
&& \quad -  \Bigl. \bigl(P_\phi^{1/3}(k_1)P_\phi^{2/3}(k_2) P_\phi(k_3)
+\mbox{(perm.)}\bigr)\Bigr]\nonumber \;.
\end{eqnarray}
and approximate the non-Gaussianity due to modification of the initial
Bunch-Davies vacuum in canonical single field action (although the
latter peaks on squashed or collinear triangles). As in the previous
example, $B_\Phi^{\rm fol}$ is suppressed in the squeezed
configurations. Unlike $B_\Phi^{\rm eq}$ however, the folded shape
induces a scale-dependent bias at large scales  (see
\S\ref{sub:dbiask}).

\subsection{Statistics of the linear mass density field}

The Bardeen's curvature potential $\Phi(\vx)$ is  related to the
linear density perturbation  $\delta_0(\vk,z)$  at redshift $z$ 
through the relation
\begin{equation}
\delta_0(\vk,z)={\cal M}(k,z)\Phi(\vk)\;,
\label{eq:poisson}
\end{equation}
where
\begin{equation}
{\cal M}(k,z)=\frac{2}{3}\frac{k^2 T(k) D(z)}{\Omega_{\rm m} H_0^2}\;.
\label{eq:transfer}
\end{equation}
Here, $T(k)$ is the matter transfer function normalized to unity as
$k\to 0$, $\Omega_{\rm m}$ is the present-day matter density, $D(z)$
is the linear growth rate normalized to $1+z$. Eq.(\ref{eq:poisson})
is important as it provides the connection between the primeval
curvature perturbations and the low redshift mass density
field. $n$-point correlator of the linear matter density field can
thus be related to those of $\Phi(\vx)$,
\begin{equation}
\la\delta_0(\vk)\cdots\delta_0(\vk_n)\ra= \Biggl(\prod_{i=1}^n{\cal
M}(k_i)\Biggr) \la\Phi(\vk_1)\cdots\Phi(\vk_n)\ra\;.
\end{equation}
Smoothing unavoidably arises when comparing observations of the large
scale structure with theoretical predictions. Perturbation theory
(PT),  which is valid only in the weakly nonlinear regime
\cite{2002PhR...367....1B}, or the spherical collapse model,  which
ignores the strongly nonlinear internal dynamics  of the collapsing
regions \cite{1972ApJ...176....1G,1996ApJS..103....1B}, require that
the small-scale nonlinear fluctuations be smoothed out.  For this
reason, we introduce the {\it smoothed} linear density field
$\delta_R$,
\begin{equation}
\delta_R(\vk,z)={\cal M}(k,z) W_R(k)\Phi(\vk)\equiv\TM(k,z)\Phi(\vk)\;,
\end{equation}
where $W_R(k)$ is a (spherically symmetric) window function of
characteristic radius $R$ or mass scale $M$. We will assume a top-hat
filter in configuration space throughout. Furthermore, since $M$ and
$R$ are equivalent variables, we shall indistinctly use the notation
$\delta_R$ and $\delta_M$ in  what follows.

\subsection{Topological defects models}

In addition to inflationary scenarios, there is a whole class of
models, known as topological defect models, in which cosmological
fluctuations are sourced by an inhomogeneously distributed component
which contributes a small fraction of the total energy momentum tensor
\cite{1999NewAR..43..111D,2000csot.book.....V}. The density field is
obtained as the convolution of a discrete set of points with a
specific density profile.  Defects are deeply rooted in particle
physics as they are expected  to form at a phase transition. Since the
early Universe may have plausibly undergone several phase transitions,
it is rather unlikely that no defects at all were formed. Furthermore,
high redshift tracers of the LSS may be superior to CMB at finding
non-Gaussianity sourced by topological defects
\cite{2001MNRAS.325..412V}. However, CMB observations already provide
stringent limits on the energy density of a defect component
\cite{2009ApJS..180..330K}, so we shall only minimally discuss the
imprint of these scenarios in the large scale structure.

\section{Evolution of the matter density field with primordial NG}
\label{sec:matterng}

In this Section, we summarize a number of results relative to the
effect of primordial NG on the mass density field. These will be
useful to understand the complexification that arises when considering
biased tracers of the density field (see \S\ref{sec:lssprobes}).

\subsection{Setting up non-Gaussian initial conditions}

Investigating the impact of non-Gaussian initial conditions (ICs) on
the large scale structure traced by galaxies etc. requires simulations
large enough so that many long wavelength modes are sampled. At the
same time, the simulations should resolve the dark matter halos
hosting the observed galaxies or quasars (QSOs), so that one can
construct halo samples whose statistical properties mimic as closely
as possible those of the real data. This favors the utilization of
pure N-body simulations, for which a larger  dynamical range can be
achieved, rather than  computationally  more expensive hydrodynamical
simulations.

The evolution of the matter density field with primordial
non-Gaussianity has been studied in series of large cosmological
N-body simulations seeded with Gaussian and non-Gaussian initial
conditions, see e.g.
\cite{1991MNRAS.248..424M,1992MNRAS.259..652W,1993MNRAS.264..749C,
1996ApJ...462L...1G,1999MNRAS.310..511W,2004MNRAS.350..287M,
2008MNRAS.390..438G,2008PhRvD..77l3514D,2009MNRAS.396...85D,
2010MNRAS.402..191P,2009arXiv0911.4768N}.  For generic non-Gaussian
(scalar) random fields, we face the problem of setting up numerical
simulations with a prescribed correlation structure
\cite{2001PASP..113.1009V}. For the equilateral and folded type of
non-Gaussianity, this task is not easily accomplished (because it
requires the calculation  of a number of convolutions which are
computationally demanding). However, for  primordial NG described by a
local mapping such as the $\floc$ model,  this is a rather
straightforward operation. This is the reason why most numerical 
studies of structure  formation with inflationary NG have so far 
implemented the local shape solely.

\subsection{Mass density probability distribution}

In the absence of primordial NG, the probability distribution function
(PDF)  of the initial smoothed density field, i.e. the probability
that a randomly placed cell of  volume $V$ has some specific density,
is Gaussian. Namely, all normalized or reduced {\it smoothed}
cumulants $S_J$ of order $J\geq 3$ are  zero. An obvious signature of
primordial NG would thus be an initially non-vanishing skewness
$S_3=\la\delta_R^3\ra_c/\la\delta_R^2\ra^2$ or kurtosis
$S_4=\la\delta_R^4\ra_c/\la\delta_R^2\ra^3-3/\la\delta_R^2\ra$
\cite{1993MNRAS.264..749C,1993ApJ...408...33L,1995MNRAS.274..730L}.
Here, the subscript $c$ denotes the connected piece of the $n$-point
moment that cannot be simplified into a sum over products of lower order
moments. At third order for instance, the cumulant of the smoothed
density field is an integral of the 3-point function,
\begin{equation}
\la\delta_R^3\ra=\int\!\!\frac{d^3 k_1}{(2\pi)^3}\int\!\!\frac{d^3 k_2}
{(2\pi)^3}\int\!\!\frac{d^3 k_3}{(2\pi)^3}\,B_R(\vk_1,\vk_2,\vk_3,z)\;,
\end{equation}
where
\begin{equation}
\fl
B_R(\vk_1,\vk_2,\vk_3,z)=\TM(k_1,z)\TM(k_2,z)\TM(k_3,z)
B_\Phi(\vk_1,\vk_2,\vk_3)
\end{equation}
is the bispectrum of the smoothed linear density fluctuations at
redshift $z$. Note that, while $S_3(R,z)\propto D(z)^{-1}$, the
product $\sigma S_3(R)$ does not depend on redshift. Over the range of
scale $0.1\lesssim R\lesssim 100\hmpc$ accessible to LSS observations,
$\sigma S_3^{(1)}\!(R)\equiv \sigma S_3(R,\fnl=1)$ is a weakly
monotonically decreasing function of $R$ that is of  magnitude  $\sim
10^{-4}$ for the local, equilateral and folded templates discussed
above.  Strictly speaking, all reduced moments should be specified to
fully characterise the density PDF, but a reasonable description of
the density distribution can be achieved with moments up to the fourth
order.

Numerical and analytic studies generally find that a density PDF
initially skewed  towards positive values produces more overdense
regions, whereas a  negatively skewed distribution produces larger
voids.  Gravitational instabilities also generate a positive skewness
in the density field, reflecting the fact that the evolved density
distribution exhibits an extended tail towards large overdensities
\cite{1980lssu.book.....P,1984ApJ...279..499F,1991MNRAS.253..727C,
1992ApJ...394L...5B,1993ApJ...412L...9J,1993ApJ...402..387L}. This
gravitationally-induced signal eventually dominates the primordial
contribution such that, at fixed normalization amplitude, non-Gaussian
scenarios deviate more strongly from the fiducial Gaussian model at
high redshift.  More precisely, the time evolution of the normalized
cumulants $S_J$ can be worked out for generic Gaussian and
non-Gaussian ICs using PT, or the simpler spherical collapse
approximation. For Gaussian ICs, PT predicts that the normalized
cumulants be time-independent to lowest non-vanishing order, with a
skewness $S_3\approx 34/7$, whereas for non-Gaussian ICs, the linear
contribution to the cumulants decays as
$S_J(R,z)=S_J(R,\infty)/D^{J-2}(z)$
\cite{1994ApJ...429...36F,1995ApJ...442...39J,1996MNRAS.279..557C,
1998MNRAS.301..524G}.

The  persistence of the primordial hierarchical amplitude
$S_J(R,\infty)$ sensitively depends upon the magnitude of $S_N$,
$N\geq J$, relative to unity. For example, an initially large
non-vanishing kurtosis could source skewness with a time-dependence
and amplitude similar to that induced by nonlinear gravitational
evolution \cite{1994ApJ...429...36F}. Although there is an infinite
class of non-Gaussian models, we can broadly divide them into weakly
and strongly non-Gaussian. In weak NG models, the primeval signal in
the normalized cumulants is rapidly obliterated by gravity-induced
non-Gaussianity. This is the case of hierarchical scaling models where
$n$-point correlation functions satisfy $\xi_n\propto \xi_2^{n-1}$
with $\xi_2\ll 1$ at large scales. By contrast, strongly NG initial
conditions dominate the evolution of the normalized cumulants. This
occurs when the hierarchy of correlation functions obeys the
dimensional scaling $\xi_n\propto \xi_2^{n/2}$, which arises in the
particular case of $\chi^2$ initial conditions
\cite{2000ApJ...542....1S} or in defect models such as texture
\cite{1991PhRvL..66.3093T,1996ApJ...462L...1G,2000PhRvD..62b1301D}.
These scaling laws have been successfully confronted with numerical
investigations of the evolution of cumulants
\cite{1996ApJ...462L...1G,1999MNRAS.310..511W}. 

Although gravitational clustering tends to erase the memory of initial
conditions, numerical simulations of non-Gaussian initial conditions
show that the occurrence of highly underdense and overdense regions is
very sensitive to the presence of primordial NG. In fact, the  imprint
of primordial NG is best preserved in the low density tail of  the PDF
$P(\rho|R)$ of the evolved density field $\rho$ smoothed at scale $R$
\cite{2008MNRAS.390..438G,2010MNRAS.402.2397L}. A satisfactory
description  of this measurement can be obtained from an Edgeworth
expansion of the initial mass density field. At high densities
$\rho\gg 1$, the non-Gaussian modification approximately scales as
$\rho^{3/5}$ whereas, at low densities $\rho\simeq 0$, the deviation
is steeper and behaves as  $\rho^{6/5}$ \cite{2009MNRAS.395.1743L}.

\subsection{Power spectrum and bispectrum}

Primordial non-Gaussianity also imprints a signature in Fourier space
statistics of the matter density field as positive values of $\fnl$
tend to increase the small scale matter power spectrum $\pmm(k)$
\cite{2004PhRvD..69j3513S,2008MNRAS.390..438G,2008PhRvD..78l3534T} and
the large scale matter bispectrum $B_\delta(\vk_1,\vk_2,\vk_3)$
\cite{2004PhRvD..69j3513S,2009PhRvD..80l3002S}.  

In the weakly nonlinear regime where 1-loop PT  applies, the Fourier
mode of the density field for growing-mode initial conditions reads
\cite{1984ApJ...279..499F,1986ApJ...311....6G}
\begin{equation}
\label{eq:d1loop}
\fl \delta(\vk,z) = \delta_0(\vk,z)+\int\!\!\frac{d^3 q_1}
{(2\pi)^3}\frac{d^3 q_2}{(2\pi)^3}\,\delta_{\rm D}(\vk-\vq_1-\vq_2)
F_2(\vq_1,\vq_2)\delta_0(\vq_1,z)\delta_0(\vq_2,z) \;.
\end{equation}
The kernel $F_2(\vk_1,\vk_2)=5/7+\mu(k_1/k_2+k_2/k_1)/2+2\mu^2/7$,
where $\mu$ is the cosine of the angle between $\vk_1$ and $\vk_2$,
describes the nonlinear 2nd order  evolution of the density field.  It
is nearly independent of $\Omega_{\rm m}$ and $\Omega_\Lambda$ and
vanishes in the (squeezed) limit $\vk_1=-\vk_2$.  At 1-loop PT, the
second term in the right-hand side of  Eq.(\ref{eq:d1loop}) generates
a correction to the mass  power spectrum,
\begin{equation}
P_\delta^{\rm NG}(k,z) = P_0(k,z)
+\left[P_{(22)}(k,z)+P_{(13)}(k,z)\right]+\Delta P_\delta^{\rm NG}(k,z)\;.
\end{equation}
Here, $P_0(k)$ is the linear matter power spectrum at redshift $z$,
$P_{(22)}$ and $P_{(13)}$ are the standard one-loop contributions in
the case of Gaussian ICs
\cite{1986ApJ...311....6G,1992PhRvD..46..585M},  and
\begin{equation}
\label{eq:p12ng}
\Delta P_\delta^{\rm NG}(k,z)=2\int\!\!\frac{d^3q}{(2\pi)^3}
F_2(\vq,\vk-\vq)B_0(-\vk,\vq,\vk-\vq,z)
\end{equation}
is the correction due to primordial NG \cite{2008PhRvD..78l3534T}.
This last term scales as $D^3(z)$ such that the effect of
non-Gaussianity is largest at low redshift. Most importantly,
$F_2(\vk_1,\vk_2)$ vanishes in the limit $\vk_1=-\vk_2$ as a
consequence of the causality of gravitational instability. This
strongly suppresses Eq.~(\ref{eq:p12ng}) at small wavenumbers, even in
the local $\floc$ model for which $B_0(-\vk,\vq,\vk-\vq)$ is maximized
in the squeezed limit $|\vk|\to 0$. For $\floc\sim {\cal O}(10^2)$,
the magnitude of the correction is at a per cent level in the weakly
nonlinear regime $k\lesssim 0.1\hmmpc$, in good agreement with
simulations
\cite{2009MNRAS.396...85D,2009arXiv0911.4768N,2009arXiv0911.0017G}.
Extensions of the renormalization group description of dark matter
clustering \cite{2007JCAP...06..026M}  to non-Gaussian initial density
and velocity perturbations can improve the  agreement up to
wavenumbers $k\lesssim 0.25\hmmpc$
\cite{2007PhRvD..76h3517I,2009arXiv0912.4276B}.

To second order in PT, the matter bispectrum $B_\delta(k_1,k_2,k_3)$
is the sum of a primordial contribution and of two terms induced by
gravitational instability
\cite{1984ApJ...279..499F,1994ApJ...426...14C} (Here and henceforth we
omit the explicit $z$-dependence for brevity),
\begin{eqnarray}
\label{eq:bispm}
\fl
B_\delta(\vk_1,\vk_2,\vk_3) = && B_0(\vk_1,\vk_2,\vk_3)
+\Bigl[2 F_2(\vk_1,\vk_2)P_0(k_1)P_0(k_2)+\mbox{(cyc.)}\Bigr]
\nonumber \\ 
&& \quad +\int\!\!\frac{d^3 q}{(2\pi)^3}\Bigl[F_2(\vq,\vk_3-\vq)
T_0(\vq,\vk_3-\vq,\vk_1,\vk_2)+\mbox{(cyc.)}\Bigr]\;,
\end{eqnarray}
where $T_0(\vk_1,\vk_2,\vk_3,\vk_4)$ is the primordial trispectrum of
the density field. Note that a similar expression can be derived for
the matter trispectrum, which turns out to be less sensitive to
gravitationally induced nonlinearities \cite{2001ApJ...553...14V}.
The reduced bispectrum $Q_3$ is conveniently defined as
\begin{equation}
Q_3(\vk_1,\vk_2,\vk_3) = \frac{B_\delta(\vk_1,\vk_2,\vk_3)}
{\Bigl[P_\delta(k_1)P_\delta(k_2)+\mbox{cyclic}\Bigr]}\;.
\end{equation}
For Gaussian initial conditions, $Q_3$ is independent of time and, at
tree-level PT, is constant and equal to $Q_3(k,k,k)=4/7$ for
equilateral configurations \cite{1984ApJ...279..499F}. For general
triangles moreover, it approximately retains this simple behavior,
with a dependence on triangle shape through $F_2(\vk_1,\vk_2)$
\cite{2004PhRvD..69j3513S}. The inclusion of 1-loop corrections
greatly  improves the agreement with the numerical data
\cite{2010arXiv1003.0007S}. An important feature property of the matter
bispectrum is the fact that the primordial part scales as $Q_3\propto
1/\TM(k)$ for approximately equilateral triangles and under the
assumption that $\floc$ is scale-independent
\cite{2004PhRvD..69j3513S}. This ``anomalous'' scaling considerably
raises the ability of the matter bispectrum to constrain primordial NG
of the local $\floc$ type. Unfortunately, neither the matter
bispectrum  nor the power spectrum are directly observable with the
large scale structure of the Universe. Temperature anisotropies in the
redshifted 21cm background from the pre-reionization epoch could in
principle furnish a direct measurement of these quantities
\cite{2004PhRvL..92u1301L,2005MNRAS.363.1049C,2007ApJ...662....1P},
but foreground contamination may severely hamper any detection.
Weak lensing is another direct probe of the dark matter, although 
we can only observe it in projection along the line of sight
\cite{2001PhR...340..291B}. 

As we will see shortly however, a similar scaling is also present in
the power spectrum and bispectrum of observable tracers of the large
scale  structure such as galaxies. It is this unique signature that
will make future all-sky LSS surveys competitive with forthcoming 
CMB experiments.

\section{LSS probe of primordial non-Gaussianity}
\label{sec:lssprobes}

Discrete and continuous tracers of the large scale structure such as
galaxies, the Ly$\alpha$ forest, the 21cm hydrogen line etc.,  provide
a distorted image of the matter density field. In CDM cosmologies,
galaxies form inside overdense regions \cite{1978MNRAS.183..341W} and
this introduces a bias between the matter and galaxy distributions
\cite{1984ApJ...284L...9K}.  As a result, distinct samples of galaxies
trace the matter distribution differently,  the most luminous galaxies
preferentially residing in the most  massive DM halos. This biasing
effect, which concerns most tracers of the large scale structure, is
still poorly understood. Models of galaxy  clustering assume for
instance that the galaxy biasing relation only depends on the local
mass density, but the actual biasing could be more complex
\cite{2008PhRvD..78j3503D, 2009JCAP...08..020M}. Because of biasing,
tracers of the large scale structure will be affected by primordial
non-Gaussianity in a different way than the mass density field. In
this Section, we describe a number of methods exploiting the abundance
and clustering properties of biased tracers to constrain the level of
primordial NG. We focus on galaxy clustering as it provides the
tightest limits on primordial NG (see \S\ref{sec:limits}).

\subsection{Halo finding algorithm}

Locating groups of bound particles, or DM halos, in simulations  is
central to the methods described below. In practice, one aims at
extracting halo catalogs with statistical properties similar to those
of observed galaxies or QSOs. This, however, proves to be quite
difficult because the relation  between observed galaxies and halos is
somewhat uncertain. Furthermore, there is freedom at defining a halo
mass.

A important ingredient is the choice of the halo identification
algorithm. There are two categories of halo finder: i) spherical
overdensity (SO) finder \cite{1994MNRAS.271..676L} with overdensity
threshold $\Delta_{\rm vir}(z)\sim 200$ and ii)  Friends-of-Friends
(FOF) finder with a linking length $b\sim 0.15-0.2$
\cite{1985ApJ...292..371D}. The mass of a SO halo is defined by the
radius at which the inner overdensity  exceeds $\Delta_{\rm vir}(z)$
times the background density  $\bar{\rho}(z)$, whereas the mass of a
FOF halo is the number of linked particles. Here, we will present
results mostly for SO halos, as their mass estimate is more closely
connected to  the predictions of the spherical collapse model, on
which most of the analytic formulae presented in this Section are
based. The question of how the spherical overdensity masses can be
mapped onto friends-of-friends masses remains a matter of debate
\cite{2009ApJ...692..217L}.   Clearly however, since the peak height
depends on the halo mass $M$ through the variance $\sigma(M)$,  any
systematic difference will be reflected in the value of $\nu$
associated to a specific halo sample.

\subsection{Abundances of voids and bound objects}
\label{sub:abundance}

A departure from Gaussianity can significantly affect the abundance of
highly biased tracers of the mass density field, as their frequency
sensitively depends upon the tails of the initial density PDF
\cite{1988ApJ...330..535L,1989ApJ...345....3C,1987MNRAS.228..407C}.
The (extended) Press-Schechter approach has been  extensively applied
to ascertain the magnitude of this effect. Because it depends only on
the skewness, it is weakly sensitive to the shape of the primordial
bispectrum.

\subsubsection{Press-Schechter approach}

The Press-Schechter theory \cite{1974ApJ...187..425P} and its 
extentions based on excursion sets
\cite{1990MNRAS.243..133P,1991ApJ...367...45C,1991ApJ...379..440B}
predict that the number density $n(M,z)$ of halos of mass $M$ at
redshift $z$ is entirely specified by a  multiplicity function
$f(\nu)$,
\begin{equation} 
n(M,z)=\frac{\bar{\rho}}{M^2}\,f(\nu)\,\frac{d\ln\nu}{d\ln M}\;,
\label{eq:fnu}
\end{equation} 
where the peak height $\nu(M,z)=\dc(z)/\sigma(M)$ is the typical
amplitude of fluctuations that produce those halos. Here and
henceforth, $\sigma(M)$ denotes the variance of the initial density
field $\delta_M$ smoothed on mass scale $M\propto R^3$ and linearly
extrapolated to present epoch, whereas $\dc(z)\approx 1.68 D(0)/D(z)$
is the critical linear overdensity for (spherical) collapse at
redshift $z$. In the standard Press-Schechter approach, $n(M,z)$ is
related to the level excursion probability $P(>\delta_c,M)$ that the
linear  density contrast of a region of mass $M$ exceeds $\delta_c(z)$,
\begin{equation}
f(\nu)= -2 \frac{\bar{\rho}}{M}\,\frac{dP}{dM} = \sqrt{\frac{2}{\pi}}
\nu\,e^{-\nu^2/2}
\end{equation}
where the last equality assumes Gaussian initial conditions. The
factor of 2 is introduced to account for the contribution of low
density regions embedded in overdensities at scale $>M$.  In the
extended Press-Schechter theory, $\delta_M$ evolves with the mass
scale $M$ and $f(\nu)$ is the probability that a trajectory is
absorbed by the constant barrier $\delta=\delta_c$ (as is appropriate
in the spherical collapse approximation) on mass scale $M$. In
general, the exact form of $f(\nu)$ depends on the barrier shape
\cite{1999MNRAS.308..119S} and the filter shape
\cite{2009arXiv0903.1249M}. Note also that  $\int\!
d\ln\nu\,f(\nu)=1$, which ensures that all the mass is contained in
halos.

Despite the fact that the Press-Schechter mass function overpredicts
(underpredicts) the abundance of low (high) mass objects, it can be
used to estimate the fractional deviation from Gaussianity. In this
formalism, the non-Gaussian fractional correction to the multiplicity
function is $R(\nu,\fnl)\equiv
f(\nu,\fnl)/f(\nu,0)=(dP/dM)(>\dc,M,\fnl)/(dP/dM)(>\dc,M,0)$, which is
readily computed  once the non-Gaussian density PDF $P(\delta_M)$ is
known.  In the simple extensions proposed by
\cite{2000ApJ...541...10M} and \cite{2008JCAP...04..014L},
$P(\delta_M)$ is expressed as the inverse transform of a cumulant
generating function.  In \cite{2008JCAP...04..014L}, the saddle-point
technique is applied directly to $P(\delta_M)$. The resulting
Edgeworth expansion is then used to obtain $P(>\dc,M)$. Neglecting
cumulants beyond the skewness, one obtain
\begin{equation}
 R_{_{\rm LV}}(\nu,\fnl) \approx  
 1+\frac{1}{6}\,\sigma S_3\left(\nu^3-3\nu\right)-
\frac{1}{6}\frac{d\left(\sigma S_3\right)}
{d\ln\nu}\left(\nu-\frac{1}{\nu}\right)
\label{eq:fnulv}
\end{equation} 
after integration over regions above  $\dc(z)$. In
\cite{2000ApJ...541...10M}, it is the level excursion probability
$P(>\dc,M)$ that is calculated within the saddle-point approximation.
This approximation asymptotes to the exact large mass tail, which
exponentially deviates from the Gaussian tail. To enforce the
normalization of the resulting mass function, one may define
$\nu_\star=\delta_\star/\sigma$ with
$\delta_\star=\dc\sqrt{1-S_3\dc/3}$, and use
\cite{2000ApJ...541...10M,2009arXiv0906.1042V}
\begin{equation}
\nu_\star f(\nu_\star)=M^2\,\frac{n_{_{\rm NG}}(M,z)}
{\bar{\rho}}\frac{d\ln M}{d\ln\nu_\star}\;.
\end{equation}
The fractional deviation from the Gaussian mass function then becomes
\begin{equation}
R_{_{\rm MVJ}}(\nu,\fnl)\approx \exp\biggl(\frac{\nu^3}{6}\sigma S_3\biggr)
\Biggl[-\frac{\sigma\nu^2}{6\nu_\star}\frac{d S_3}{d\ln\nu}+
\frac{\nu_\star}{\nu}\Biggr]\;.
\label{eq:fnumvj}
\end{equation}
Both formulae have been shown to give reasonable agreement with
numerical simulations of non-Gaussian cosmologies
\cite{2007MNRAS.382.1261G,2009MNRAS.396...85D,2009MNRAS.398..321G}
(but note that \cite{2007MNRAS.376..343K,2008PhRvD..77l3514D} have
reached somewhat different conclusions).  Expanding
$\delta_\star=\delta_c\sqrt{1-S_3\delta_c/3}$ at the first order shows
that these two theoretical expectations differ in the coefficient of
the $\nu\sigma S_3$ term. Therefore, it is also instructive to
consider the approximation \cite{2010PhRvD..81b3006D}
\begin{equation}
R(\nu,\fnl) \approx \exp\biggl(\frac{\nu^3}{6}\sigma S_3\biggr)
\Biggl[1-\frac{\nu}{2}\sigma S_3-\frac{\nu}{6}\frac{d(\sigma S_3)}
{d\ln\nu}\Biggr]\;,
\label{eq:fnuthiswork}
\end{equation}
which is designed to match better the Edgeworth expansion of
\cite{2008JCAP...04..014L} when the peak height is $\nu\sim 1$.   Note
that, when the primordial trispectrum is large (which, in the local
model, would happen if $\gloc\sim 10^6$), terms involving the kurtosis
should also be  included
\cite{2000ApJ...541...10M,2008JCAP...04..014L,
2010PhRvD..81b3006D,2009arXiv0910.5125M}. In this case, it is also
important to take into account a possible renormalization of the
fluctuation amplitude, $\sigma_8\to\sigma_8+\delta\sigma_8$, to which
the high mass tail of the mass function is exponentially sensitive
\cite{2010PhRvD..81b3006D}. Finally, note also that
\cite{2008PhRvD..77l3514D,2010MNRAS.402..191P} parametrize the
fractional correction using N-body  simulations.

\begin{figure}
\center \resizebox{0.6\textwidth}{!}{\includegraphics{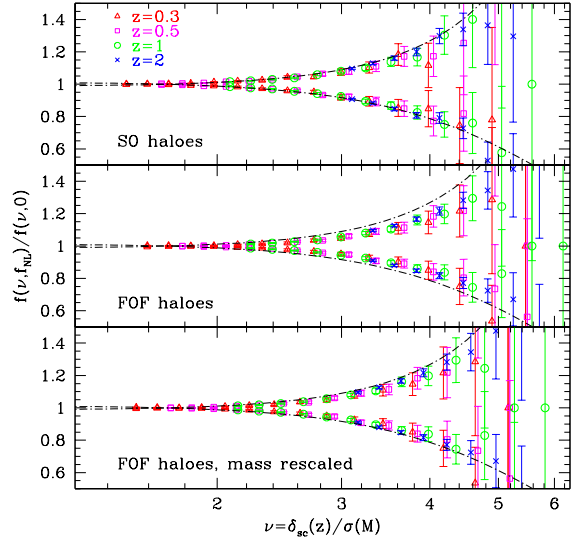}}
\caption{Fractional deviation from the Gaussian mass function as a
function of the peak height $\nu=\dc/\sigma$. Different  symbols refer
to different redshifts as indicated. The solid curve is the
theoretical prediction Eq.~(\ref{eq:fnuthiswork}) at $z=0$ based on
an Edgeworth expansion of the dark matter probability distribution
function. In the top panel, halos are identified using a  spherical
overdensity (SO) finder with a redshift-dependent overdensity
threshold $\Delta_{\rm vir}(z)$.  In the middle panel, a
Friends-of-Friends (FOF) finding algorithm with linking length $b=0.2$
is used. The bottom panel shows the effect of decreasing the FOF mass
by 20\% (see text).  In all panels, error bars denote Poisson
errors. For illustration, $M=10^{15}\hmsun$ corresponds to $\nu=3.2$,
5.2, 7.7 at redshift $z=0$, 1 and 2, respectively. Similarly,
$M=10^{14}\hmsun$ and $10^{13}\hmsun$ correspond to $\nu=1.9$, 3, 4.5
and 1.2, 1.9, 2.9 respectively.}
\label{fig:fnu}
\end{figure}

Figure \ref{fig:fnu} shows the effect of primordial NG of the local
$\floc$ type on the halo mass function. The dotted-dashed curve
represents the approximation Eq.(\ref{eq:fnuthiswork}). While the
agreement is reasonable for the SO halos (top panel), the theory
significantly overestimates the deviation measured in the FOF mass
function with linking length $b=0.2$ (middle panel). A similar effect
is noted in \cite{2009MNRAS.398..321G}, who makes the replacement
$\delta_c\to\delta_c\sqrt{q}$ with $q\approx 0.75$ to fit their
measurement of $R(\nu,\floc)$ based on FOF halos. References
\cite{2009arXiv0903.1250M,2009arXiv0903.1251M}  provide a physical
motivation of this replacement by demonstrating that the diffusive
nature of the collapse barrier introduces a similar  factor. However,
an overlooked but important fact is that the FOF and SO mass estimates
systematically deviate from each other. In Fig.\ref{fig:fnu} in
particular, the FOF mass is on average 20\% larger than the SO
mass. As shown in the bottom panel of Fig.~\ref{fig:fnu}, rescaling
the FOF mass to account for this difference removes most of the
discrepancy with the FOF data. This illustrates an important point:
the impact of primordial NG on the statistics of DM halos is sensitive
to systematics caused by the choice of the halo finder. As we will see
below, this may also be true for the non-Gaussian halo bias.

More sophisticated formulations based on extended Press-Schechter (EPS)
theory and/or modifications of the collapse criterion look promising
since they can reasonably reproduce  both the Gaussian halo counts and
the dependence on $\fnl$
\cite{2009arXiv0903.1250M,2009MNRAS.398.2143L,2009MNRAS.399.1482L}.
The probability of first upcrossing can, in principle, be derived for
any non-Gaussian density field and any choice of smoothing filter
\cite{2000MNRAS.314..354A,2002ApJ...574....9I}.  For a general filter,
the non-Markovian dynamics generates additional terms in the
non-Gaussian correction to the mass function that arise from 3-point
correlators of the  smoothed density $\delta_M$ at different mass
scales \cite{2009arXiv0903.1250M}. However, large error bars still
make difficult to test for the presence of such sub-leading terms. For
generic moving barriers $B(\sigma)$ such as those appearing in models
of triaxial collapse \cite{2001MNRAS.323....1S,2008MNRAS.388..638D},
the leading contribution to the non-Gaussian correction approximately
is \cite{2009MNRAS.398.2143L}
\begin{equation}
R(\nu,\fnl)\approx  1+\frac{1}{6}\,\sigma
S_3\,H_3\Bigl(\frac{B(\sigma)}{\sigma}\Bigr)\;,
\end{equation}
where $H_3(\nu)\equiv \nu^3-3\nu$, and agrees well with the observed
deviation \cite{2009MNRAS.399.1482L}.

\subsubsection{Clusters abundance}

Rich clusters of galaxies trace the rare, high-density peaks in  the
initial conditions and thus offer the best probe of the high mass tail
of the multiplicity function. To infer the cluster mass function,  the
X-ray and millimeter windows are better suited than the optical-wave
range because selection effects can be understood better. 

Following early theoretical \cite{1986ApJ...310L..21M,
1988ApJ...330..535L,1988PhRvL..61..267C,1989ApJ...345....3C,
1989A&A...215...17B} and numerical
\cite{1991MNRAS.253...35M,1991ApJ...372L..53P,1992MNRAS.259..652W,
1994MNRAS.266..524B} work on the effect of non-Gaussian initial
conditions on the multiplicity function of cosmic structures, the
abundance of clusters and X-ray counts in non-Gaussian cosmologies has
received much attention in the literature. At fixed normalization of
the observed abundance of local clusters, the proto-clusters
associated with high redshift ($2<z<4$) Ly$\alpha$ emitters are much
more likely to develop in strongly non-Gaussian models than in the
Gaussian paradigm
\cite{2004MNRAS.350..287M,2004MNRAS.353..681M,2007MNRAS.376..343K}.
Considering the redshift evolution of cluster abundances can thus
break the degeneracy between the initial density PDF and the
background cosmology. In this regards, simple extensions of  the
Press-Schechter formalism (similar to those considered above) have
been shown to capture reasonably well the cluster mass function over a
wide range of redshift for various non-Gaussian scenarios
\cite{2000MNRAS.311..781R}.   Scaling relations between the cluster
mass, X-ray temperature and Compton $y$-parameter calibrated using
theory, observations and detailed simulations of cluster formation
\cite{2006ApJ...648..956S,2006ApJ...650..538N}, can then be exploited
to predict the observed distribution functions of X-ray and SZ signals
and assess the capability of cluster surveys to test the nature of the
initial conditions
\cite{1998ApJ...494..479C,1999MNRAS.310.1111K,2000ApJ...532....1R,
2002MNRAS.331...71B,2006MNRAS.368.1583S,2007MNRAS.380..637S,
2007ApJ...658..669S,2009MNRAS.397.1125F,2009arXiv0909.4714R}.

An important limitation of this method is that, for a realistic amount
of primordial NG, the non-Gaussian signal imprinted in cluster
abundances is small compared to the systematics plaguing current and
upcoming surveys
\cite{1997A&A...320..365O,2000ApJ...534..565H,2004MNRAS.351..375A}.
Given the current uncertainties in the redshift evolution of clusters
(one commonly assumes that clusters are observed at the epoch they
collapse \cite{2000ApJ...534..565H}), the selection effects in the
calibration of X-ray and SZ fluxes with  halo mass, the freedom in the
definition of the halo mass, the degeneracy with the normalization
amplitude $\sigma_8$   (for positive $\fnl$, the mass function is more
enhanced at the high  mass end, and this  is similar to an increase in
the amplitude of  fluctuations $\sigma_8$ \cite{2007JCAP...06..024M})
and the low number statistics, the prospects  of  using the cluster
mass function only to place competitive limits on $\fnl$ with the
current data are small.  A two-fold improvement in cluster mass
calibration is required to provide constraints comparable to CMB
measurements \cite{2004MNRAS.351..375A}.

\subsubsection{Voids abundance}

The frequency of cosmic voids, which is strongly sensitive to the  low
density tail of the initial mass PDF, offers another probe of
non-Gaussian initial conditions \cite{2009JCAP...01..010K}. The
Press-Schechter formalism can also be applied to ascertain the
magnitude of this effect. Voids are defined as regions of mass $M$
whose density is less than some critical value $\delta_v\leq 0$ or,
alternatively, as regions for which the three eigenvalues of the tidal
tensor \cite{1970Afz.....6..581D} lie below some critical value
$\lambda_v\leq 0$
\cite{2009JCAP...01..010K,2009MNRAS.395.1743L,2009ApJ...701L..25S,
2009MNRAS.399.1482L}. An important aspect in the calculation of the
mass function of voids is the over-counting of voids located inside
collapsing regions. This voids-in-clouds problem, as identified by
\cite{2004MNRAS.350..517S}), can be solved within the excursion set
theory by studying a two barriers problem: $\dc$ for halos and
$\delta_v$ for voids. Including this effect reduces the frequency of
the smallest voids \cite{2009MNRAS.399.1482L}. Neglecting this
complication notwithstanding, the differential number density of voids
of radius $R$ is \cite{2009JCAP...01..010K,2009ApJ...701L..25S}
\begin{equation}
\frac{dn}{dR} \approx \frac{9}{2\pi^2}\sqrt{\frac{\pi}{2}}\,
\frac{|\nu_v|}{R^4}\, e^{-\nu_v^2/2}\, \frac{d\ln|\nu_v|}{d\ln M}
\Biggl[1-\frac{1}{6}\sigma S_3\,H_3\Bigl(|\nu_v|\Bigr)\Biggr]\;,
\end{equation}
where $\nu_v=\delta_v/\sigma_M$.  While a positive $\fnl$ produces
more massive halos, it generates fewer large voids
\cite{2009JCAP...01..010K,2009MNRAS.399.1482L}.  Hence, the effect is
qualitatively different from a simple rescaling of the normalization
amplitude $\sigma_8$. A joint analysis of both abundances of clusters
and cosmic voids might thus provide interesting constraints on the
shape of the primordial 3-point function. There are, however, several
caveats to this method, including the fact that there is no unique way
to define voids \cite{2009JCAP...01..010K,2008ApJ...684L...1C}.
Clearly, voids identification algorithms will have to be tested on
numerical simulations \cite{2008MNRAS.387..933C} before a robust
method can be applied to real data.

\subsection{Galaxy 2-point correlation}
\label{sub:dbiask}

In Gaussian cosmologies,   correlations  of galaxies and clusters can
be amplified relative to the mass distribution 
\cite{1984ApJ...284L...9K}.  Before this was realized, it was argued
that primeval fluctuations need to be non-Gaussian
\cite{1980ApJ...236..351D,1983ApJ...274....1P}  to explain the
observed strong correlation of Abell clusters
\cite{1983ApJ...270...20B,1983SvAL....9...41K}. Along these lines,
\cite{1986ApJ...310...19G} pointed out that  primordial
non-Gaussianity could significantly increase the amplitude  of the
two-point correlation of galaxies and clusters on large scales,
However, except from \cite{2006A&A...457..385A} who showed that
correlations of high density peaks in non-Gaussian models are
significantly stronger than in the Gaussian model with identical mass
power spectrum,  subsequent work focused mostly on abundances
(\S\ref{sub:abundance}) or higher order statistics such as  the
bispectrum  (\S\ref{sub:bispgal}). It is only recently that
\cite{2008PhRvD..77l3514D} have demonstrated the strong
scale-dependent bias arising in non-Gaussian models of the local
$\floc$ type.

\subsubsection{The non-Gaussian bias}

In the original derivation of \cite{2008PhRvD..77l3514D}, the
Laplacian is applied to the left and right hand side of
$\Phi=\phi+\floc\phi^2$ to show that, upon substitution of the Poisson
equation, the overdensity in the neighborhood of density peaks is
spatially modulated by a factor proportional to the local value of
$\phi$. Taking into account the coherent motions induced by
gravitational instabilities, the scale-dependent bias correction reads
\begin{equation}
\Delta b_\kappa(k,\floc)= 3\floc \bigl[b_1(M)-1\bigr]\dc(0)
\frac{\Omega_{\rm m}H_0^2}{k^2 T(k) D(z)}\;,
\label{eq:dbiask}
\end{equation}
where $b_1(M)$ is the linear, Gaussian halo bias.  The original result
of \cite{2008PhRvD..77l3514D} missed out a multiplicative factor of
$T(k)^{-1}$ which was introduced subsequently by
\cite{2008ApJ...677L..77M} upon a derivation of Eq.~(\ref{eq:dbiask})
in the limit of high density peaks. The peak-background split approach
\cite{1986ApJ...304...15B,1989MNRAS.237.1127C,1999MNRAS.308..119S}
promoted by \cite{2008JCAP...08..031S} shows that the scale-dependent
bias applies to any tracer of the matter density field whose
(Gaussian)  multiplicity function depends on the local mass density
only. In this approach, the Gaussian piece of the potential is
decomposed into short- and long-wavelength modes,
$\phi=\phi_l+\phi_s$.  This provides an intuitive explanation of the
effect in terms of a local rescaling of the small-scale amplitude of
matter fluctuations by a factor $1+2\floc\phi_l$ (see also
\cite{2008PhRvD..77l3514D,2008PhRvD..78l3507A,2009arXiv0911.0017G}).
As emphasized in \cite{2008PhRvD..77l3514D}, the scale-dependence
arises from the fact that the non-Gaussian curvature perturbations
$\Phi(\vx)$ are related to density fluctuations through the Poisson
equation (\ref{eq:poisson}). There is no such effect in the $\chi^2$
model \cite{1999ApJ...510..531P,2000MNRAS.313..141V} nor in
texture-seeded cosmologies \cite{1991ApJ...383....1C} for instance.

The derivation of \cite{2008ApJ...677L..77M}, based on the clustering
of regions of the smoothed density field $\delta_M$ above threshold
$\delta_c(z)$, is formally valid for  high density peaks
only. However, it is general enough to apply to  any shape of
primordial bispectrum \cite{2009ApJ...706L..91V}. In the high
threshold limit $\nu\gg 1$,   the 2-point correlation function of the
level excursion set  can be expressed as \cite{1986ApJ...310L..21M}
\begin{equation}
\label{eq:corrlevel}
\fl \xi_{>\nu}(\vr) = -1 +
  \exp\Biggl\{\sum_{n=2}^\infty\sum_{j=1}^{n-1}\frac{\nu^n\sigma^{-n}}
  {j!(n-j)!}\xi_R^{(n)}\!\left(\begin{array}{cc}\vx_1,\cdots,\vx_1,  &
  \vx_2,\cdots,\vx_2 \\ j~\mbox{times} &
  (n-j)~\mbox{times}\end{array},z=0 \right) \Biggr\}  \nonumber \;,
\end{equation}
where $\vr=\vx_1-\vx_2$. For most non-Gaussian models in which the
primordial 3-point function is the dominant correction, this expansion
can be truncated at the third order and Fourier transformed to yield
the non-Gaussian correction $\Delta P_{>\nu}(k)$ to the power
spectrum. Assuming a small level of primordial NG, we can also write
$\Delta P_{>\nu}(k)\approx 2 b_{\rm L}\Delta b_\kappa P_R(k)$, where
$b_{\rm L}=b_1(M)-1\approx \nu^2/\dc$ is the Lagrangian bias, and
eventually obtain
\begin{equation}
\label{eq:dbfnl}
\Delta b_\kappa(k,\fnl)\equiv b_\phi(k){\cal F}(k,\fnl)=
\Biggl(\frac{2 b_{\rm L}\dc(z)}{\TM(k,0)}\Biggr){\cal F}(k,\fnl)\;.
\end{equation}
The dependence on the shape of the 3-point function is encoded in the
function ${\cal F}(k,\fnl)$ \cite{2008ApJ...677L..77M,2009ApJ...706L..91V},
\begin{equation}
\label{eq:fk}
\fl
{\cal F}(k,\fnl) = \frac{1}{16\pi^2\sigma^2}\int_0^\infty\!\! 
dk_1\,k_1^2 \TM(k_1,0)\int_{-1}^{+1}\!\!d\mu\,\TM(\sqrt{\alpha},0)
\frac{B_\Phi(k_1,\sqrt{\alpha},k)}{P_\Phi(k)} \nonumber \;,
\end{equation}
where $\alpha^2=k^2+k_1^2+2\mu k k_1$. Note that, for $\fnl<0$, this
fist order approximation always breaks down at sufficiently small $k$
because $\Delta P_{>\nu}(k)<0$.

The scale-dependent bias induced by the equilateral and folded
bispectrum shape is computed in \cite{2009ApJ...706L..91V}.  To get
insights into the large scale behavior of $\Delta b_\kappa(k,\fnl)$,
let us identify the dominant contribution to ${\cal F}(k,\fnl)$ in the
limit $k\ll 1$.  Setting $\TM(\sqrt{\alpha},0)\approx\TM(k_1,0)$  and
expanding $P_\phi(\sqrt{\alpha})$ at second order in the ratio
$k/k_1$, we arrive at  
\numparts
\begin{eqnarray}
\label{eq:bapprox}
\fl
{\cal F}(k,\floc) \approx \floc \\
\fl
{\cal F}(k,\feq) \approx \feq\,\Bigl[3\,\Sigma_{_{R}}\!
\Bigl(\frac{2(n_s-4)}{3}\Bigr)k^{\frac{2(4-n_s)}{3}}
+ \frac{1}{2}\left(n_s-4\right)\Sigma_{_{R}}\!\left(-2\right)
k^2\Bigr]\,\sigma^{-2} \\
\fl
{\cal F}(k,\ffol) \approx \frac{3}{2}\ffol\,
\Sigma_{_{R}}\!\Bigl(\frac{n_s-4}{3}\Bigr)k^{\frac{4-n_s}{3}}\,\sigma^{-2}\;, 
\end{eqnarray}
\endnumparts
assuming no running scalar index, i.e. $dn_s/d\ln k=0$. The auxiliary 
function $\Sigma_{_{R}}\!(n)$ is defined as
\begin{equation}
\Sigma_{_{R}}\!(n)=\frac{1}{2\pi^2}\int_0^\infty\!\!dk\,k^{(2+n)}\,\TM(k,0)^2
P_\phi(k)\;. 
\end{equation}
Hence, we have $\Sigma_{_{R}}\!(0)\equiv \sigma^2$. For a nearly
scale-invariant spectrum $n_s\approx 1$, we obtain ${\cal
F}(k,\ffol)\propto k$  and ${\cal F}(k,\feq)\propto k^2$, such that
the non-Gaussian bias is  $\Delta b_\kappa\propto k^{-1}$ and  $\Delta
b_\kappa=$const. for the folded and equilateral bispectrum,
respectively. Therefore, at large scales, the scale-dependence of the
non-Gaussian bias is much smaller for the folded template, and nearly
absent for the equilateral shape. This make them much more difficult
to detect with galaxy surveys \cite{2009ApJ...706L..91V}. However,
the equilateral and folded non-Gaussian bias depend sensitively upon
the mass scale $M$ through the multiplicative factor
$\sigma^{-2}$. For example, choosing $R=5\hmpc$ instead of  $R=1\hmpc$
would increase the effect by a factor of $\sim 3$. In the high peak
limit, $\sigma^{-2}\approx b_{\rm L}/\dc(z)$,  which cancels out the
dependence on redshift but enhances the sensitivity  to the halo bias,
i.e. $\Delta b_\kappa\propto b_{\rm L}^2$. By contrast,  $\Delta
b_\kappa\propto b_{\rm L}$ for the local $\floc$ model.

At this point, it is appropriate to mention a few caveats to these
calculations. Firstly, Eq. (\ref{eq:dbfnl}) assumes that the tracers
form after a spherical collapse, which may be a good approximation for
the massive halos only. If one instead considers the ellipsoidal
collapse dynamics, in which the evolution of a perturbation  depends
upon the three eigenvalues of the initial tidal shear, $\dc(0)$ should
be replaced by its ellipsoidal counterparts $\dec(0)$ which  is always
larger than the spherical value \cite{2001MNRAS.323....1S}. In this
model, the scale-dependent bias $\Delta b_\kappa$ is thus enhanced by
a factor $\dec(0)/\dc(0)$ \cite{2008PhRvD..78l3507A}. Secondly,
Eq.~(\ref{eq:dbfnl})  assumes that the biasing of the surveyed objects
is described by the peak height $\nu$ only, or equivalently, the
hosting halo mass $M$. However, this may not be true for quasars whose
activity may be triggered by merger of halos
\cite{2003MNRAS.343..692H,2005ApJ...630..705H}. Reference
\cite{2008JCAP...08..031S} used the EPS formalism to estimate the bias
correction $\Delta b_{\rm merger}$ induced by recent mergers,
\begin{equation}
\Delta b_{\rm merger}=\dc^{-1}\;,
\end{equation}
so the factor $b_1(M)-1$ should be replaced by
$b_1(M)-1-\dc^{-1}\approx b_1(M)-1.6$. The validity of this result
should  be evaluated with cosmological simulations of quasars
formation. In this respect, semi-analytic models of galaxy formation
suggest that merger-triggered objects such as quasars do not cluster
much differently than other tracers of the same mass
\cite{2009arXiv0909.0003B}. However, this does not mean that the same
should hold for the non-gaussian scale dependent bias. Still, since
the recent merger model is an extreme case it seems likely that the
actual bias correction is $0<\Delta b_{\rm merger}<\dc^{-1}$.
Thirdly, the scale-dependent bias has been derived using the Newtonian
approximation to the Poisson equation, so one may wonder whether
general relativistic (GR) corrections to $\TM(k)^{-1}$ suppress the
effect on scales comparable to the Hubble radius. Reference
\cite{2005JCAP...10..010B} showed how large scale primordial NG
induced by GR corrections propagates onto small scales once
cosmological perturbations reenter the Hubble radius in the matter
dominated era. This effect generates a scale-dependent bias
comparable, albeit of opposite sign to that induced by local NG
\cite{2009ApJ...706L..91V}. However, this issue deserves further
clarification as \cite{2009PhRvD..79l3507W} have recently argued that
there are no GR corrections to the non-Gaussian bias and that the
scaling $\Delta b_\kappa\propto k^{-2}$ applies down to the smallest
wavenumbers.

Finally, we can also ask ourselves whether  higher-order terms in the
series expansion (\ref{eq:corrlevel})  furnish corrections to the
non-Gaussian bias of magnitude comparable to Eq.(\ref{eq:dbiask}).  In
the $\floc$ model, the power spectrum of biased tracers of the density
field can also be obtained from a local Taylor series in the evolved
(Eulerian) density contrast $\delta$ and the Gaussian part $\phi$ of
the initial (Lagrangian) curvature perturbation
\cite{2008PhRvD..78l3519M,2009arXiv0911.0017G}. Using this approach,
it can be shown that the halo power spectrum arising from the first
order terms of the local bias expansion can be cast into the form
\cite{2008PhRvD..78l3519M}
\begin{equation}
\label{eq:phhloc}
\phh(k)=\bigl[ b_1(M) + \floc b_\phi(k) \bigr]^2 P_R(k)
\end{equation}
Hence, we obtain a second order term proportional to
$(\floc)^2\TM^{-2}P_R(k)=(\floc)^2P_\phi(k)$ which, however,
contributes only at very small wavenumber $k\lesssim 0.001\hmpc$.
There is another second order correction to the halo power spectrum
that reads \cite{2010PhRvD..81b3006D}
\begin{equation}
\Delta\phh(k)=\frac{4}{3}(\floc)^2 \Bigl[b_1(M)-1\Bigr]^2
\dc^2(z)\,S_3^{(1)}\!(M) \TM(k,0) P_\phi(k)\;.
\end{equation}
Its magnitude relative to the term linear in $\floc$,
Eq.(\ref{eq:dbiask}),  is approximately $0.03$ at redshift $z=1.8$ and
for a halo mass $M=10^{13}\hmsun$.  Although this contribution becomes
increasingly important at higher redshift, it is fairly small for
realistic values of $\floc$. All this suggests that
Eq.~(\ref{eq:dbiask}) is the dominant contribution to the non-Gaussian
bias in the wavenumber range $0.001\lesssim k\lesssim 0.1\hmmpc$.

\subsubsection{Comparison with simulations}

In order to fully exploit the potential of forthcoming large scale
surveys, a number of studies have tested the theoretical prediction
Eq.(\ref{eq:dbiask}) against the outcome of large numerical
simulations
\cite{2008PhRvD..77l3514D,2009MNRAS.396...85D,2010MNRAS.402..191P,
2009MNRAS.398..321G,2009arXiv0911.4768N,2009arXiv0911.0017G}.

At the lowest order, there are two additional albeit relatively
smaller corrections to the Gaussian bias which arise from the
dependence of both the halo number density $n(M,z)$ and the matter
power spectrum $\pmm(k,z)$ on primordial NG \cite{2009MNRAS.396...85D}.
Firstly, assuming the peak-background split holds, the change in  the
mean number density of halos induces a scale-independent offset which
we denote $\Delta b_{\rm I}(\floc)$.  In terms of the non-Gaussian
fractional correction $R(\nu,\floc)$ to  the mass function, this
contribution is
\begin{equation}
\label{eq:dbiasi}
\Delta b_{\rm I}(\floc) =-\frac{1}{\sigma}\frac{\partial}{\partial\nu}
\ln\Bigl[R(\nu,\floc)\Bigr]\;.
\end{equation}
It is worth noticing that $\Delta b_{\rm I}(\floc)$ has a sign
opposite to that of $\floc$, because the  bias decreases when the mass
function goes up. Secondly, we also need to account for the change in
the matter power  spectrum (see \S\ref{sec:matterng}). As a result,
non-Gaussianity of the $\floc$ type adds a correction $\Delta
b(k,\floc)$  to the bias $b(k)$ of dark matter halos that reads
\cite{2009MNRAS.396...85D}
\begin{equation}
\fl
\Delta b(k,\floc)=\Delta b_\kappa(k,\floc)+\Delta b_{\rm I}(\floc)
+b_1(M)\left(\frac{\pmm(k,\floc)}{\pmm(k,0)}-1\right)
\label{eq:dbias}
\end{equation}
at first order in $\floc$. The linear bias $b_1$ needs to be measured
accurately as it  controls the strength of the scale-dependent bias
correction $\Delta b_\kappa$. In this respect, the ratio
$\pmh(k)/\pmm(k)$, where $\pmh(k)$ is the halo-matter cross power
spectrum, is  a better proxy for the halo bias as it is less sensitive
to shot-noise. 

Reference \cite{2009MNRAS.396...85D} find that the inclusion of these
extra terms substantially improves the comparison between the theory
and the simulations.  Considering only the scale-dependent shift
$\Delta b_\kappa$ leads to an apparent suppression of the effect in
simulations relative to the theory.  Including the scale-independent
offset $\Delta b_{\rm I}$ considerably improves the agreement at all scales.
Finally, adding the
scale-dependent term $b_1(M)(P_{\rm mm}(k,\floc)/P_{\rm mm}(k,0)-1)$
further adjusts  the match at small scale $k\gtrsim 0.05\hmmpc$ by
making the non-Gaussian  bias shift less negative. Taking into account
second- and  higher-order corrections could extend the validity of the
theory up to scales $k\sim 0.1-0.3\hmmpc$ \cite{2009arXiv0911.0017G}.

\begin{figure}
\center \resizebox{0.6\textwidth}{!}{\includegraphics{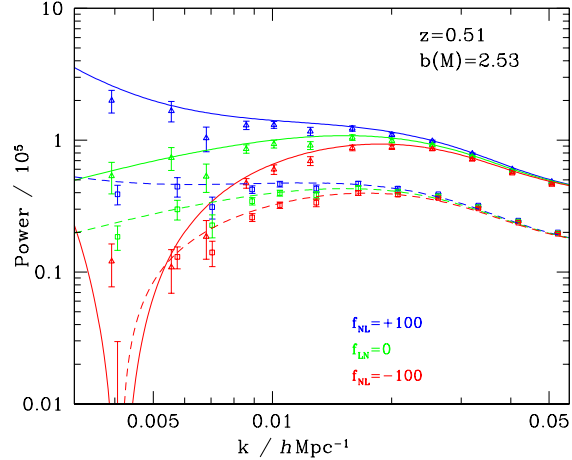}}
\caption{Halo-halo and halo-matter power spectra $\phh(k)$ and
$\pmh(k)$ measured in simulations of the Gaussian model and of the
local $\floc$ type with $\floc=\pm 100$. Halos of mass  $M>2\times
10^{13}\hmsun$ were identified at redshift $z=2$ with a  SO
finder. The linear Gaussian bias of this sample is $b_1(M)=2.53$.  The
error bars represent the scatter among 8 realizations. The solid and
dashed curve show the theoretical $\phh(k)$ and $\pmh(k)$  obtained
wih the non-Gaussian bias correction Eq.(\ref{eq:dbias}).  For
$\floc=-100$, the cross-power spectrum is negative on scales
$k\lesssim 0.005\hmmpc$, in good agreement with the theoretical
prediction.}
\label{fig:powng}
\end{figure}

Auto- and cross-power analyses may not agree with each other if the
halos and dark matter do not trace each other on scale  $k\lesssim
0.01\hmmpc$ where the non-Gaussian bias is large,  i.e. if there is
stochasticity.  Fig.\ref{fig:powng} shows $\pmh(k)$ and $\phh(k)$
averaged over 8 realisations of the models with $\floc=0,\pm 100$. The
same Gaussian random seed field $\phi$ was used in each set of runs so
as to minimize the sampling variance. Measurements of the non-Gaussian
bias correction  obtained with the halo-halo or the halo-matter power
spectrum are in a good agreement with each other, indicating that
non-Gaussianity does not induce  stochasticity and the predicted
scaling Eq.(\ref{eq:dbiask}) applies equally  well for the auto- and
cross-power spectrum. However, while a number of numerical studies of
the $\floc$ model have confirmed the scaling  $\Delta
b_\kappa(k,\floc)\propto\TM(k)^{-1}$ and the redshift dependence
$\propto D(z)^{-1}$
\cite{2008PhRvD..77l3514D,2009MNRAS.396...85D,2010MNRAS.402..191P,
2009MNRAS.398..321G}, the exact amplitude of the non-Gaussian bias
correction remains somewhat debatable. Reference
\cite{2009MNRAS.396...85D}, who use SO halos and include the
scale-independent offset  $\Delta b_{\rm I}$, find satisfactory
agreement with the theory. By contrast,
\cite{2009MNRAS.398..321G,2010MNRAS.402..191P}, who use FOF halos,
find that Eq.(\ref{eq:dbiask}) is a good fit to the simulations only
upon replacing $\dc$ by $q\dc$ with $q\simeq 0.75$. Part of the
discrepancy may be due to the fact that
\cite{2009MNRAS.398..321G,2010MNRAS.402..191P} do not include $\Delta
b_{\rm I}$, which leads to an apparent suppression of the effect.
Another possible source of discrepancy may be choice of the halo
finder. In this regards, Fig.~\ref{fig:dbfof} shows the non-Gaussian
bias correction obtained with FOF halos. The best-fit values of
$\floc$ are somewhat below the input values of $\pm 100$, in agreement
with the findings of
\cite{2009MNRAS.398..321G,2010MNRAS.402..191P}. This indicates that
the choice of halo finder also affects the  magnitude of the
non-Gaussian halo bias. Discrepancies have also been observed between
the theoretical and measured non-Gaussian bias corrections  in
non-Gaussian models of the local cubic-order coupling $\gloc\phi^3$
\cite{2010PhRvD..81b3006D}. Understanding these results will clearly
require a better theoretical modeling of halo clustering.

\subsubsection{Redshift distortions}

Peculiar velocities generate systematic differences between the
spatial distribution of data in real and redshift space. These
redshift  distortions must be properly taken into account in order to
extract  $\fnl$ from redshift surveys.  On the linear scales of
interest, the redshift space power spectrum of biased tracers reads as
\cite{1987MNRAS.227....1K,1998ASSL..231..185H}
\begin{equation}
P^s(k,\mu)=\Bigl[b_1^2 P_\delta(k)+2 b_1 f\mu^2P_{\delta\theta}(k)
+f^2\mu^4P_\theta(k)\Bigr]\;,
\end{equation}
where $P_{\delta\theta}$ and $P_\theta$ are the density-velocity and
velocity divergence power spectra, $\mu$ is the cosine of the angle
between the wavemode $\vk$ and the line of sight and $f$ is the
logarithmic derivative of the growth factor. For $P_\theta$, the
1-loop correction due to primordial NG is identical to
Eq.(\ref{eq:p12ng}) provided  $F_2(\vk_1,\vk_2)$ is replaced by the
kernel  $G_2(\vk_1,\vk_2)=3/7+\mu(k_1/k_2+k_2/k_1)/2+4\mu^2/7$
describing the 2nd order evolution of the velocity divergence
\cite{2000ApJ...542....1S}.  For $P_{\delta\theta}$, this correction is
\begin{equation}
\fl
\Delta P_{\delta\theta}^{\rm NG}(k)=\int\!\!\frac{d^3\vq}{(2\pi)^3}\,
\Bigl[F_2(\vq,\vk-\vq)+G_2(\vq,\vk-\vq)\Bigr] B_0(-\vk,\vq,\vk-\vq)\;.
\end{equation}
Again, causality implies that $G_2(\vk_1,\vk_2)$ vanishes in the limit
$\vk_1=-\vk_2$. For unbiased tracers with $b_1=1$, the  linear Kaiser
relation is thus recovered at large scales $k\lesssim 0.01\hmmpc$ (see
also \cite{2010MNRAS.402.2397L}). For biased tracers, we still expect
the Kaiser formula to be valid, but the distortion parameter $\beta$
should now be equal to  $\beta=f/(b_1+\Delta b_\kappa)$, where $\Delta
b_\kappa(k,\fnl)$ is the  scale-dependent bias induced by the
primordial non-Gaussianity.

\subsubsection{Mitigating cosmic variance and shot-noise}

Because of the finite number of large scale wavemodes accessible to a
survey, any large scale measurement of the power spectrum is limited
by the cosmic (or sampling) variance caused by the random  nature of
the wavemodes. For discrete tracers such as galaxies, the shot noise
is another source of error. For weak primordial NG, the relative error
on the power spectrum $P$ is $\sigma_P/P\approx
1/\sqrt{N}(1+\sigma_n^2/P)$, where $N$ is the number of independent
modes measured and $\sigma_n^2$ is the shot-noise
\cite{1994ApJ...426...23F}. Under the standard assumption of Poisson
sampling, $\sigma_n^2$ equals the inverse of the number density
$1/\bar{n}$ and causes  a scale-independent enhancement of the power
spectrum. The extent to which one can improve the observational limits
on the nonlinear will strongly depend on our ability to minimize the
impact  of these two sources of errors. By comparing differently
biased tracers  of the same surveyed volume
\cite{2009PhRvL.102b1302S,2009JCAP...10..007M} and suitably weighting
galaxies (by the mass of their host halo for instance)
\cite{2009JCAP...03..004S,2009PhRvL.103i1303S}, it should be possible
to circumvent these problems and considerably improve the detection
level.

\begin{figure}
\center \resizebox{0.7\textwidth}{!}{\includegraphics{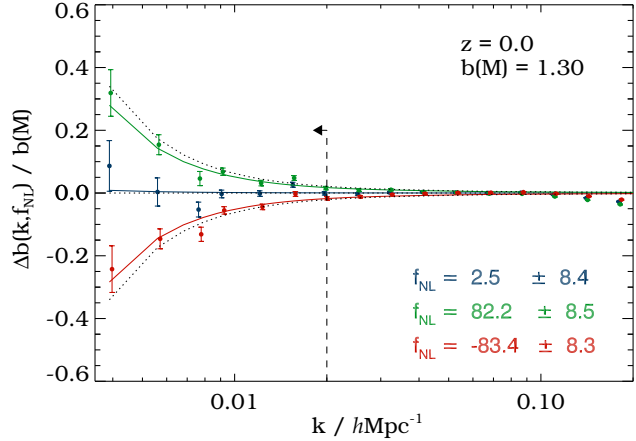}}
\caption{Fractional correction to the Gaussian halo bias in the
$\floc=\pm 100$ and Gaussian models. In constrast to
Fig.~\ref{fig:powng}, halos were identified with a FOF finder of
linking length  $b=0.2$. Only the wavemodes to the left of the
vertical line were used to fit $\Delta b_\kappa(k,\floc)$. The
best-fit value of $\floc$ and the corresponding 1$\sigma$ error is
quoted for each model (Figure taken from \cite{nicohamaus}).}
\label{fig:dbfof}
\end{figure}

Figure \ref{fig:dbfof} illustrates how the impact of sampling variance
on the measurement of $\floc$ can be mitigated. Namely, the data
points show the result of taking the ratio
$\phh(k,\floc)/\pmm(k,\floc)$  for each set of runs with same Gaussian
random seed field $\phi$ before averaging over the realisations. This
procedure is  equivalent to the multi-tracers method advocated by
\cite{2009PhRvL.102b1302S}. Here, $\pmm$ can be thought as mimicking
the power spectrum of a nearly unbiased tracer of the mass density
field with high number density. Although, in practical applications,
using the dark matter field works better \cite{nicohamaus}, in real
data  $\pmm$ should be replaced by a tracer of the same surveyed
volume  different than the one used to compute $\phh$.  Figure
\ref{fig:dbfof} also shows that, upon taking out most of the cosmic
variance, there is some residual noise caused by the discrete nature
of the dark matter halos. As shown recently \cite{2009PhRvL.103i1303S}
however, weighting the halos according to their mass can dramatically
reduce the shot noise relative to the Poisson  expectation, at least
when compared against the  dark matter. Applying such a weighting may
thus significantly improve the error on the nonlinear parameter
$\floc$, but this should be explored in realistic simulations of
galaxies, especially because the halo mass $M$ may not be easily
measurable from real data \cite{nicohamaus}.  This approach
undoubtedly deserves further attention as it has the potential to
substantially improve the extraction of the primordial non-Gaussian
signal from galaxy surveys.

To conclude this Section, we note that, while the PDF of power values
$P(\vk)$ has little discriminatory power (for large surveyed volume,
it converges towards the Rayleigh distribution as a consequence of the
central limit theorem) \cite{1995PhRvD..51.6714F}, the covariance of
power spectrum measurements (which is sensitive to the selection
function, but also to correlations among the phase of the Fourier
modes) may provide quantitative limits on certain type of non-Gaussian
models \cite{1994ApJ...426...23F,1996MNRAS.283L..99S}.

\subsection{Galaxy bispectrum and higher order statistics}
\label{sub:bispgal}

Higher statistics of biased tracers, such as the galaxy bispectrum,
are of great interest as they are much more sensitive to the shape of
the primordial 3-point function than the power spectrum
\cite{2004PhRvD..69j3513S,2007PhRvD..76h3004S,2009ApJ...703.1230J,
2009PhRvD..80l3002S,2009arXiv0911.4768N}. Therefore, they could break
some of the degeneracies affecting the non-Gaussian halo bias (For
example, the leading order scale-dependent correction to the Gaussian
bias induced by the local quadratic and cubic coupling are fully
degenerate \cite{2010PhRvD..81b3006D}).

\subsubsection{Normalized cumulants of the galaxy distribution}

The skewness of the galaxy count probability distribution function
could provide constraints on the amount of non-Gaussianity in the
initial conditions. As discussed in \S\ref{sec:matterng} however, it
is difficult to disentangle the primordial and gravitational causes of
skewness in low redshift data unless the initial density field is
strongly non-Gaussian. The first analyzes of galaxy catalogs in terms
of count-in-cells densities all reached the conclusion that the
skewness (and higher-order moments) of the observed galaxy count PDF
is consistent with the value induced by gravitational instabilities of
initially Gaussian fluctuations
\cite{1991MNRAS.253..727C,1992ApJ...398L..17G,1993ApJ...403..450G,
1993ApJ...417...36B,1994ApJ...429...36F,1995MNRAS.274.1049B}. Back
then however, most  of the galaxy samples available were not large
enough to accurately determine the cumulants $S_J$ at large scales
\cite{1995ApJ...443..469L}. Despite the 2 orders of magnitude increase
in surveyed volume, these measurements are still sensitive to cosmic
variance, i.e. to the presence of massive super-clusters or large
voids. Nevertheless, the best estimates of the first normalized
cumulants $S_J$ of the galaxy PDF strongly suggest that high order
galaxy correlation functions follow the hierarchical scaling predicted
by the gravitational clustering of Gaussian ICs
\cite{2004MNRAS.352.1232C}. There is no evidence for strong
non-Gaussianity in the initial density field as might by seeded by
cosmic strings or textures \cite{2006MNRAS.373..759F}.

The genus statistics of constant density surfaces through the  galaxy
distribution measures the relative abundance of low and high density
regions as a function of the smoothing scale $R$  and, therefore,
could also be used as a diagnostic tool for primordial
non-Gaussianity. While for a Gaussian random field the genus curve
(i.e. the genus number as a function of the density contrast) is
symmetric about $\delta_R=0$ regardless the value of $R$, primordial
NG and nonlinear gravitational evolution can disrupt this symmetry
\cite{1996ApJ...460...51M}. The effect of non-Gaussian ICs on the
topology of the galaxy distribution has been explored in a number of
papers
\cite{1992MNRAS.259..652W,1993MNRAS.260..572C,1996ApJ...463..409M,
2001ApJ...556..641H,2008MNRAS.385.1613H}. For large values of $R$ and
a realistic amount of primordial NG, the genus statistics can also be
expanded in a series whose coefficients are the normalized cumulants
$S_J$ of the smoothed galaxy density field. Therefore, the genus
statistics essentially provides another measurement of the (large 
scale) cumulants of the galaxy distribution
\cite{2008ApJ...675...16G,2009MNRAS.394..454J}.

\subsubsection{Galaxy bispectrum}

Most of the scale-dependence of the primordial $n$-point functions is
integrated out in the normalized cumulants, which makes them weakly
sensitive to primordial NG. However, while the effect of non-Gaussian
initial conditions, galaxy bias, gravitational instabilities etc. are
strongly degenerated in the $S_J$, they imprint distinct signatures in
the galaxy bispectrum $B_{\rm h}(\vk_1,\vk_2,\vk_3)$, an accurate
measurement of which could thus constrain the shape of the primordial
3-point function.

In the original derivation of \cite{2007PhRvD..76h3004S}, the large
scale (unfiltered) galaxy bispectrum in the $\floc$ model is given by
\begin{eqnarray}
\label{eq:bispg}
\fl B_{\rm h}(\vk_1,\vk_2,\vk_3) = && b_1^3 B_0(\vk_1,\vk_2,\vk_3) 
+ b_1^2 b_2\Bigl[P_0(k_1) P_0(k_2)+\mbox{(cyc.)}\Bigr]  \nonumber \\ && 
\quad + 2b_1^3\Bigl[F_2(\vk_1,\vk_2)P_0(k_1)P_0(k_2) +\mbox{(cyc.)}
\Bigr]\;.
\end{eqnarray}
Here, $b_1$ and $b_2$ are the first- and second-order bias parameters
that describe galaxy biasing relation assumed local and deterministic
\cite{1993ApJ...413..447F}. The first term in  the right-hand side is
the primordial contribution which, for  equilateral configurations and
in the $\floc$ model, scales as $\TM(k,z)^{-1}$ like in the matter
bispectrum, Eq.(\ref{eq:bispm}). The two last terms are the
contribution from nonlinear bias and the tree-level correction  from
gravitational instabilities, respectively. They have the smallest
signal in squeezed configurations.

As recognized by \cite{2009ApJ...703.1230J,2009PhRvD..80l3002S},
Eq.(\ref{eq:bispg}) misses an important term that may significantly
enhance the sensitivity of the galaxy bispectrum to non-Gaussian
initial conditions. This contribution is sourced by the trispectrum
$T_R(\vk_1,\vk_2,\vk_3,\vk_4)$ of the smoothed mass density field,
\begin{equation}
\frac{1}{2}b_1^2 b_2\int\!\!\frac{d^3 q}{(2\pi)^3}\,
T_R(\vk_1,\vk_2,\vq,\vk_3-\vq)+\mbox{(2 perms.)}\;,
\end{equation}
and reduces at large scales to the sum of the linearly evolved
primordial trispectrum $T_0(\vk_1,\vk_2,\vk_3,\vk_4)$ and a coupling
between the primordial bispectrum $B_0(\vk_1,\vk_2,\vk_3)$  (linear in
$\fnl$) and the second order PT corrections (through the kernel
$F_2(\vk_1,\vk_2)$). In the case of local non-Gaussianity and for
equilateral configurations, the first piece proportional to $T_0$
scales as $(\floc)^2 k^{-4}$ times the Gaussian tree-level prediction,
with the same redshift dependence. Hence, it  is similar to the second
order correction $(\floc)^2{\cal M}_R^{-2}P_R(k)$ that appears in the
halo power spectrum (see Eq.\ref{eq:phhloc}). The second piece linear
in $\fnl$ generates a signal at large scales  for essentially all
triangle shapes in the local model as well as in the case of
equilateral NG.  This second contribution is maximized in the squeezed
limit (where it is one order of magnitude larger than the result
obtained by \cite{2007PhRvD..76h3004S}) which helps disentangling it
from the Gaussian terms. Note that a strong dependence on triangle
shape is also present in other NG scenarios such as the $\chi^2$ model
\cite{2000ApJ...542....1S}.

This newly derived contributions are claimed to lead to more than one
order of magnitude improvement in certain limits
\cite{2009ApJ...703.1230J}, but it is not yet clear whether these
gains can be realized in any realistic survey.  To accurately predict
the constraints that could be achieved with future measurements of the
galaxy bispectrum, a comparison of these predictions with the halo
bispectrum extracted from numerical simulations is highly
desirable. To date, the only numerical study
\cite{2009arXiv0911.4768N} has measured the halo bispectrum for some
isosceles triangles ($k_1=k_2$). While the shape dependence is in
reasonable agreement with the theory, the  observed $k$-dependence
appears to depart from the predicted scaling.

\subsection{Intergalactic medium and the Ly$\alpha$ forest}

\begin{figure}
\center
\resizebox{0.7\textwidth}{!}{\includegraphics[angle=-90]{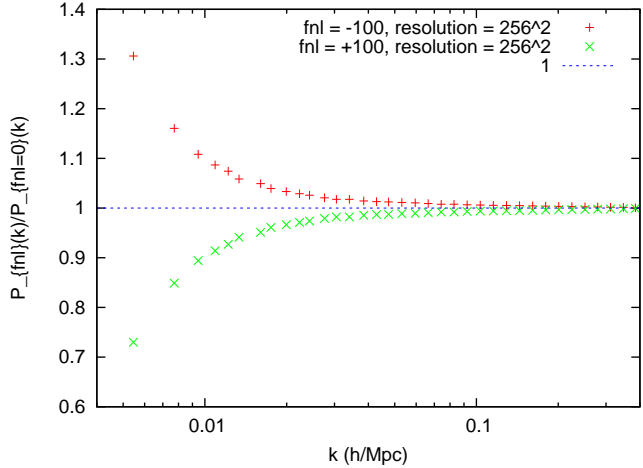}}
\caption{Ratio between the Lyman-$\alpha$ 3D flux power spectrum
extracted from simulations of Gaussian and non-Gaussian initial
conditions  at redshift $z=2$. The mean transmission is set to
$\bar{F}=0.8$ (Figure taken from \cite{shirleyho}).}
\label{fig:flux}
\end{figure}

Primordial non-Gaussianity also affects the intergalactic medium (IGM)
as a positive $\fnl$ enhances the formation of high-mass halos  at
early times and, therefore, accelerate reionization
\cite{2003MNRAS.346L..31C,2006MNRAS.371.1755A,2009MNRAS.394..133C}.
At lower redshift, small box hydrodynamical simulations of the
Ly$\alpha$ forest indicate  that non-Gaussian initial conditions could
leave a detectable signature in the Ly$\alpha$ flux PDF, power
spectrum and bispectrum  \cite{2009MNRAS.393..774V}. However, while
differences appear quite pronounced in the high transmissivity tail of
the flux PDF (i.e. in underdense regions), the Ly$\alpha$ 1D flux
power spectrum seems little affected.  Given the small box size of
these hydrodynamical simulations,  it is worth exploring the effect in
large N-body cosmological simulations using a semi-analytic modeling
of the Ly$\alpha$ forest \cite{shirleyho}.  Figure ~\ref{fig:flux}
shows the imprint of local type NG on the Ly$\alpha$ 3D flux power
spectrum (which is not affected by projection effects) extracted from
a series of large simulations  at $z=2$. The Ly$\alpha$ transmitted
flux is calculated in the Gunn-Peterson approximation
\cite{1965ApJ...142.1633G}.  A clear signature similar to the
non-Gaussian halo bias can be seen and, as expected, it is of opposite
sign since the Ly$\alpha$ forest is anti-biased relative to the mass
density field (overdensities are mapped onto relatively low flux
transmission).  A detection of this effect, although challenging in
particular because of continuum uncertainties, could be feasible with
future data sets. The Ly$\alpha$ could thus provide interesting
information on the non-Gaussian signal over a range of scale and
redshift not easily accessible to galaxy and CMB observations
\cite{2009MNRAS.393..774V,shirleyho}.

\section{Current limits and prospects}
\label{sec:limits}

As the importance of primordial non-Gaussianity relative to the
non-Gaussianity induced by gravitational clustering and galaxy bias
increases towards high redshift, the optimal strategy to constrain 
the nonlinear coupling parameter(s) with LSS is to use large scale,
high-redshift observations \cite{2001MNRAS.325..412V}.

\subsection{Existing constraints on primordial NG}

The non-Gaussian halo bias presently is the only LSS method that
provides a robust limit on the magnitude of a primordial 3-point
function of the local shape. It is a broadband effect that can be
easily measured with photometric redshifts. The authors of
\cite{2008JCAP...08..031S} have applied Eq.(\ref{eq:dbiask}) to
constrain the value of $\floc$ using a compilation of large-scale
clustering data. Their constraint arise mostly from the QSO sample  at
median redshift $z=1.8$, which covers a large comoving volume  and is
highly biased, $b_1=2.7$. They obtain
\begin{equation}
-29 < \floc < +69
\end{equation}
at 95\% confidence level.  These limits are competitive with those
from CMB measurements,  $-10<\floc<+74$ \cite{2010arXiv1001.4538K}.
It is straightforward  to translate this  2-$\sigma$ limit into a
constraint on the cubic order coupling $\gloc$ since the non-Gaussian
scale-dependent bias  $\Delta b_\kappa(k,\gloc)$ has the same
functional form as  $\Delta b_\kappa(k,\floc)$
\cite{2010PhRvD..81b3006D}. Assuming $\floc=0$, one  obtains
\begin{equation}
-3.5\times 10^5 < \gloc < +8.2 \times 10^5\;.
\end{equation}
These limits are comparable with those inferred from an analysis of
CMB data using $n$-point distribution functions,
$-5.6\times 10^5<\gloc<6.4\times 10^5$ \cite{2009arXiv0910.3196V}.

Measurements of the galaxy bispectrum in several redshift catalogs
have shown evidence for a configuration shape dependence in agreement
with that predicted from gravitational instability, ruling out
$\chi^2$ initial conditions at the 95\%
C.L. \cite{1999ApJ...521L..83F,2001ApJ...546..652S}.  Recent analyses
of the SDSS LRGs catalogue indicate that the shape dependence  of the
reduced 3-point correlation $Q_3\sim\xi_3/(\xi_2)^2$ is  also
consistent with Gaussian ICs  \cite{2007MNRAS.378.1196K}, although  a
primordial (hierarchical)  non-Gaussian contribution in the range
$Q_3\sim 0.5-3$ cannot be ruled out  \cite{2009MNRAS.399..801G}. Other
LSS probes of primordial non-Gaussianity, such as the abundance of
massive clusters, are still too affected by systematics to furnish
tight constraints on the shape and magnitude of a primordial 3-point
function, although the observation of a handful of unexpectedly
massive high-redshift clusters has been interpreted as evidence of a
substantial  degree of primordial NG
\cite{2000ApJ...530...80W,2009PhRvD..80l7302J, 2010arXiv1003.0841S}.

\subsection{Future prospects}

Improving the current limits will further constrain the physical 
mechanisms for the generation of cosmological perturbations. 

The non-Gaussian halo bias also leaves a signature in
cross-correlation  statistics of weak cosmic shear (galaxy-galaxy and
galaxy-CMB) \cite{2009PhRvD..80l3527J,2009arXiv0912.4112F} and in the
integrated Sachs-Wolfe (ISW) effect
\cite{2008JCAP...08..031S,2008PhRvD..78l3507A}.  Measurements of the
lensing bispectrum could also  constrain a number of non-Gaussian
models \cite{2004MNRAS.348..897T}.   However, galaxy clustering will
undoubtedly offer the most promising LSS diagnostic of primordial
non-Gaussianity. The detectability of a local primordial bispectrum
has been assessed in a series of papers. It is expected that future
all-sky galaxy surveys will achieve constraints of the order of
$\Delta\floc\sim 1$ assuming all systematics are reasonably under
control
\cite{2008JCAP...04..014L,2008JCAP...08..031S,2008PhRvD..78l3507A,
2008PhRvD..78l3519M,2008ApJ...684L...1C,2010RAA....10..107G,
2009JCAP...12..022S,2010arXiv1003.0456C}.  Realistic models of cubic
type non-Gaussianity \cite{2010PhRvD..81b3006D}, modifications of the
initial vacuum state or horizon-scale GR corrections
\cite{2009ApJ...706L..91V} should also be tested with future measurement
of the galaxy power spectrum.

Upcoming observations of high redshift clusters will provide increased
leverage on measurement of primordial non-Gaussianity with abundances
and possibly put limits on any nonlinear parameter $\fnl$ at the level
of a few tens \cite{2009MNRAS.397.1125F}.  Combining the information
provided by the evolution of the mass function and power spectrum of
galaxy clusters can yield constraints with a precision
$\Delta\floc\sim 10$ for a wide field survey covering half of the sky
\cite{2010arXiv1003.0841S}. Alternatively, using the full covariance
of cluster counts (which is sensitive to the non-Gaussian halo bias)
can furnish constraints of $\Delta\floc\sim 1-5$ for a Dark Energy
Survey-type experiment \cite{2009PhRvL.102u1301O,2010arXiv1003.2416C}.

As emphasized in \S\ref{sec:lssprobes} however, the exact magnitude of
the non-Gaussian halo bias is still uncertain at the $\sim$20\% level,
partly due to the freedom at the definition of the halo mass.
Understanding this type of systematics will be crucial to set reliable
constraints on a primordial non-Gaussian component.  To fully exploit
the potential of future galaxy surveys, it will also be essential to
extend the theoretical and numerical analyses to other bispectrum
shapes than the local template used so far. Ultimately, the gain that
can  be achieved will critically depend on our ability to minimize the
impact of sampling variance and shot-noise. In this regards,
multi-tracers methods combined with optimal weighting schemes should
deserve further  attention as they hold the promise to become the most
accurate method to extract the primordial non-Gaussian signal from
galaxy surveys \cite{2009PhRvL.102b1302S,2009JCAP...03..004S,
2009JCAP...10..007M,2009PhRvL.103i1303S}.

\section{Acknowledgments}

Special thanks to Nico Hamaus and Shirley Ho for sharing with us
material prior to publication. We would like  to thank Martin Crocce,
Nico Hamaus, Christopher Hirata, Shirley Ho, Ilian Iliev, Tsz Yan Lam,
Patrick McDonald, Nikhil Padmanabhan, Emiliano Sefusatti, Ravi Sheth
and Anze Slosar for their collaboration on these issues; and Tobias
Baldauf for comments on an early version of this manuscript.  This
work was supported by the Swiss National Foundation (Contract
No. 200021-116696/1) and made extensive use of the NASA Astrophysics
Data System and the arXiv.org preprint server.

\bibliographystyle{unsrt}
\bibliography{cqg_reviewng}

\end{document}